# Analogies of structural instabilities in EuTiO$_3$ and SrTiO$_3$

A. Bussmann-Holder, J. Köhler, R. K. Kremer, and J. M. Law

Max-Planck-Institute for Solid State Research, Heisenbergstr. 1, D-70569 Stuttgart, Germany

Specific heat measurements and theoretical calculations reveal an intimate analogy between EuTiO$_3$ and SrTiO$_3$. For EuTiO$_3$ a hitherto unknown specific heat anomaly is discovered at $T_A=282(1)$K which is analogous to the well known specific heat anomaly of SrTiO$_3$ at $T_A=105$K caused by an antiferrodistortive transition. Since the zone center soft phonon mode observed in both systems can be modeled with the <u>same parameters</u> we ascribe the new 282(1)K instability of EuTiO$_3$ to an antiferrodistortive phase transition. The higher transition temperature of EuTiO$_3$ as compared to SrTiO$_3$ results from spin phonon coupling.

Pacs-Index 63.20.-e, 75.80.+q

ATiO$_3$ (A=Ba, Pb, Zr, Sr, Ca, Cd, Eu) perovskites are well known for their tendencies towards numerous instabilities. While the Ba, Pb, Cd titanate compounds undergo ferroelectric phase transitions, the corresponding perovskite PbZrO$_3$ exhibits an antiferroelectric phase transition. CaTiO$_3$ (CTO) [1], SrTiO$_3$ (STO) [2 – 4], and EuTiO$_3$ (ETO) [5, 6] show pronounced long wave length optic mode softening over a large temperature range, but never become ferroelectric since quantum fluctuations suppress a long-range instability [7]. STO and CTO instead undergo an antiferrodistortive zone boundary related structural phase transition to a tetragonal phase, at $T_A=105$K [8, 9] and $T_A=837$K [10, 11], respectively, which is accompanied by an extremely small lattice distortion in STO where the c/a lattice constant ratio changes by less than 1‰ and a more pronounced one in CTO [10]. Due to the small change of c/a in STO it was difficult to reveal this phase transition and only from electron spin resonance (ESR) [12], electron paramagnetic resonance (EPR) [8, 13], and inelastic neutron scattering (INS) data [14 – 16] could it be clearly detected. ETO on the other hand, becomes antiferromagnetic (AFM) at $T_N$=5.5K [17] thereby influencing strongly the soft optic mode which abruptly increases in energy at $T_N$ [5, 6, 18]. This latter effect has been speculated to originate from multi-ferroic behavior which has substantially increased the interest in this compound. However, since a true ferroelectric instability is inhibited by quantum fluctuations, the term multi-ferroic is misleading.

While the dynamic properties of STO are well understood, namely originating from Ti d O p charge transfer [19], those of ETO are still under discussion. In STO, both, the soft zone boundary and the soft zone center mode have been shown to be caused by the configurational instability of the $O^{2-}$ ion [20] which is unstable as a free entity [21]. In a crystal a partial stabilization of the $O^{2-}$ ion is achieved through the interaction with the surrounding ions, but the basic tendency towards delocalization of the $2p^6$ electrons remains. This behavior has been termed *dynamical covalency* and been modeled within the polarizability model [19, 22 – 24] which is based on a double-well potential in the local core-shell coupling constant at the oxygen ion site. Since ETO has a variety of properties in common with STO, namely optic mode softening, suppression of a ferroelectric instability by quantum fluctuations, induced ferroelectricity in strained films [25 – 27], identical lattice constants, and identical ionic radii of Sr and Eu, it is suggestive that the dynamics can be modeled within the same theoretical approach which has already been proven successful for STO. In order to model the AFM state at low temperatures the polarizability model is extended by a spin-spin and a spin phonon interaction term which closely resembles the one studied by Jacobsen and Stevens [28] except for the use of the polarizability coordinate. The Hamiltonian is given by:

$$H = \sum_{i=1,2}\sum_n \left\{ \frac{p_{i,n}^2}{m_i} + \frac{f'}{2}(u_{1,n+1}-u_{1,n-1})^2 + \frac{1}{2}g_T w^2 + \frac{f}{2}[(u_{2,n}-u_{1,n}-w_{1,n})^2 + (u_{2,n-1}-u_{1,n}-w_{1,n})^2] \right\}$$
$$+ \sum_n \left\{ \hbar\varepsilon(u_{2,n+1}-u_{2,n})S_x^{(2,n+1)}S_x^{(2,n)} + g\beta H S_z^{(2,n)} \right\}$$

(1)

with $\beta = eh/2mc$ and $g_T = g_2 + 3g_4 <w^2>_T$ where $g_2$ is attractive while $g_4$ is the fourth order repulsive anharmonic coupling term in the polarizability coordinate $w$ which is treated in a cumulant expansion using the self-consistent phonon approximation (SPA). $u_{in}$, $m_i$ ($i=1,2$) are displacement coordinates and masses of ions $i$, respectively, where $m_1$ refers to the polarizable cluster mass $TiO_3$ and $m_2$ is the rigid ion mass of Eu. $f, f'$ are nearest and second nearest neighbor harmonic coupling constants. It is important to note that these coupling constants are *the same* as those derived for STO and only the A site sublattice mass has been changed to conform to the higher Eu mass. The coupling between the spins and the lattice, $\varepsilon$, is bilinear with respect to the A sublattice while the coupling between the polarization and the spin includes a third order coupling term according to $S_x = \omega_0 \varepsilon <S_z>/(\omega_0^2 - \omega^2)(w\{1+g_T/2f\}+u)/\cos qa$, which introduces higher order couplings due to $g_T$ analogous to Ref. 29. $\varepsilon$ modifies the $xy$ components of the $g$ tensors through the lattice oscillations, and varies linearly with the magnetic field $H$. $S_x^{(n)}, S_z^{(n)}$ are the $x, z$ components of the spin at site $n$ of the Eu atom. By introducing the definitions:

$$\frac{fg_T}{2f+g_T}=\tilde{f}, \quad \frac{4f'}{m_1}\sin^2 qa+\frac{2\tilde{f}}{m_1}=\omega_1^2, \quad \frac{4ff}{g_T m_2}\sin^2 qa+\frac{2\tilde{f}}{m_2}=\omega_2^2, \quad \omega_0^2=\frac{g\beta H}{\hbar},$$

the corresponding dispersion relations are given by:

$$(\omega^2-\omega_1^2)(\omega^2-\omega_2^2)(\omega^2-\omega_0^2)-\omega_0^2\varepsilon^2<S_z>(\omega^2-\omega_1^2)-\frac{4\tilde{f}^2}{m_1 m_2}\cos^2 qa(\omega^2-\omega_0^2)=0$$

(2)

The temperature dependence of $g_T$, where $g_2, g_4$ are determined self-consistently within the SPA, defines the soft mode temperature dependence. For small spin-lattice coupling the zero momentum optic mode softens with decreasing temperature as shown in Fig. 1 wherein the calculated soft mode frequency is compared to the experimental data of Refs. 5, 6 evidencing excellent agreement with the experiment. In this limit the coupled eq. 2 can approximately be decoupled in the long wave length limit to $\omega_F^2(q=0)\approx 2\tilde{f}/\mu$, with $\mu$ being the reduced cell mass and $\omega^2=\omega_0^2$. Obviously, the soft mode has the same temperature dependence as in the uncoupled case which applies to STO. The saturation regime of $\omega_f^2$ at temperatures T<30K is a consequence of quantum fluctuations analogous to STO.

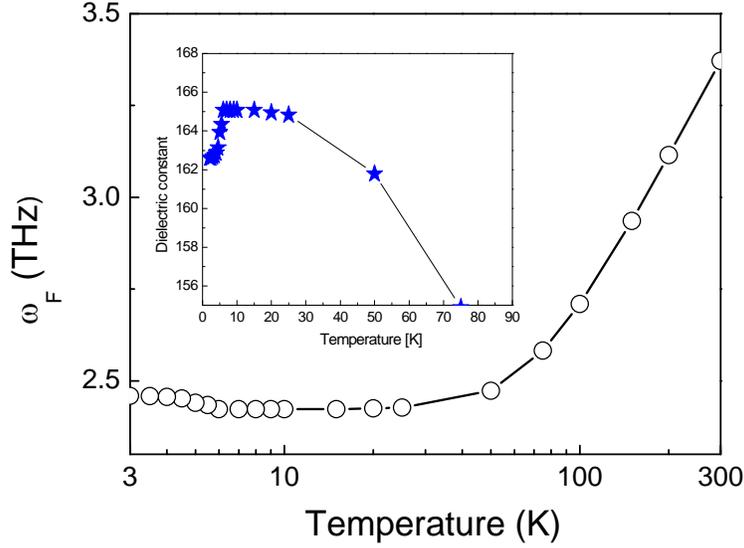

**Figure 1** Temperature dependence of the soft optic mode $\omega_F$ of ETO in a semilogarithmic plot. The full stars are experimental data from Refs. 5, 6. The inset shows the related dielectric constant. Solid lines are guides to the eye.

This regime changes if the spin-lattice coupling strength $\varepsilon \approx H$ is switched on and long-range AFM order sets in below $T_N$. Again an approximate analytic solution to eq. 2 for the soft mode exists in this limit which for large coupling strength, i.e., large fields, is given by: $\omega_F^2 = \widetilde{f}/\mu + \omega_0^2 + \omega_0 \varepsilon \sqrt{|<S_z>|}$. In accordance with experimental data [5, 6] the spin-phonon coupling depresses the dielectric constant (Fig. 1, inset) and causes an anomaly at $T_N$ as seen experimentally. Also, already at high temperatures mode-mode coupling sets in which induces a lowering of the zone boundary acoustic mode energy [29].

The dynamical properties of ETO are well modeled by the *same* parameters as used for STO. This agreement indicates that a zone boundary related phase transition as realized in STO at $T_A=105K$ should be present also in ETO. For STO it has recently been shown that this instability arises from the same polarizability effects as the zone center soft mode [20] with the transition temperature $T_A$ being given by:

$$k_B T_A = 1/3 g_4 [-g_2 + \frac{4ff}{2f'+f}] \sum_{q,j=0-2} \frac{w^2(q)}{\omega_j^2(q)}, \qquad (3)$$

where $w^2$ is the polarizability coordinate eigenvector and $\omega_j$ is defined by eq. 2. For this reason an analogous calculation has been carried through for ETO with the distinction that the second nearest neighbor interaction is more attractive than in STO due to the spin-phonon coupling term. From this calculation an antiferrodistortive phase transition is predicted to occur at T≈298K. It is important to emphasize that similar strong spin-phonon coupling has been observed in the rare earth manganites [30 – 33] where even a hybridized soft mode magnon excitation has been detected by inelastic neutron scattering [34].

Before proceeding with the experimental results it is important to mention that the self-consistently derived double-well potentials differ distinctively between STO and ETO (Fig. 2). While in STO the potential is broad with a shallow minimum, it is narrow and deep in ETO. This finding implies that the STO dynamics are more on the displacive side, whereas in ETO order / disorder type dynamics are realized which has already been addressed in Refs. [35 – 37]. By producing mixed STO-ETO crystals an interesting crossover between both should occur.

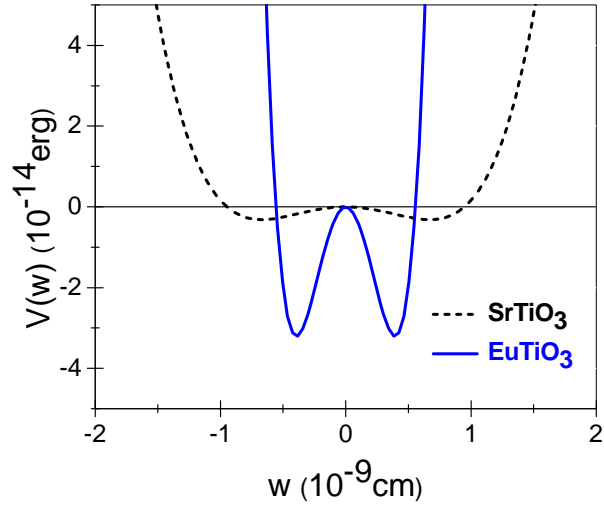

**Figure 2** The local double-well potential of STO (black line) and the one of ETO (blue line). The model parameters of ETO are the same as those of STO [28] with a mass enhancement factor of 1.73 applied to $m_2$ to account for the heavier Eu sublattice and new self-consistently derived double-well defining parameters, namely $g_2 = -41.3806, g_4 = 133.5556$.

In order to test the above prediction, ETO samples have been prepared by carefully mixing dried $Eu_2O_3$ ((Alfa, 99.99%) with $Ti_2O_3$ powder (Alfa, 99.99%), in a 1:1 ratio in an agate mortar under Ar. Then the powder was pressed to a pellet, and heated in a corundum tube under Ar for 4 d at 1400°C. The ETO sample was dark grey with a cubic lattice constant of 390.6(1) pm at room temperature according to X-ray powder diffraction data. A further temperature dependent X-ray diffraction scan did not reveal any deviations from cubic symmetry, thus seemingly disproving the expected existence of a phase transition. However, similar experience has been made with STO, where initially only ESR, EPR, and INS [8, 12 – 16] were able to see the phase transition. Later also specific heat measurements detected a tiny anomaly at $T_A$ in STO [38 – 41]. For this reason specific heat measurements (relaxation type calorimeter PPMS, Quantum Design) have first been repeated for STO (commercially available samples) and then been carried through for ETO (Fig. 3). In STO the structural instability causes an obvious anomaly in the specific heat at 105K (see inset b) to Fig. 3) which is seen in the present experiment much clearer than in the previous ones [38 – 41]. In ETO an anomaly very similar in shape and magnitude to that of STO is seen at 282(1)K, close to the theoretically expected phase transition temperature. In addition, also the transition to the AFM state is evident in the specific heat data as a λ-type anomaly (see inset a) to Fig. 3).

In order to exclude that the phase transition at 282(1)K is related to some magnetic ordering stemming from the Eu spins, the magnetic susceptibility has been carefully measured and found to be in best agreement with existing data [17, 42, 43]. While at $T_N$=5.5K clearly a deviation from linearity is seen corresponding to the AFM transition temperature, deviations from the Curie law occur at $T_A$=282(1)K could not be detected. Therefore, it is concluded that the specific heat anomaly at this temperature stems from a structural phase transition where the analogies between STO and ETO suggest that this is of antiferrodistortive origin with extremely small changes in the cubic axis relation which obscures its detection by (IR) infrared or Raman scattering techniques. In order to further substantiate this conclusion it is proposed to perform INS experiments on ETO where the modeling predicts an acoustic mode boundary softening. An additional support is also expected to come from ESR, EPR and Mössbauer measurements where a line splitting should appear at 282(1)K.

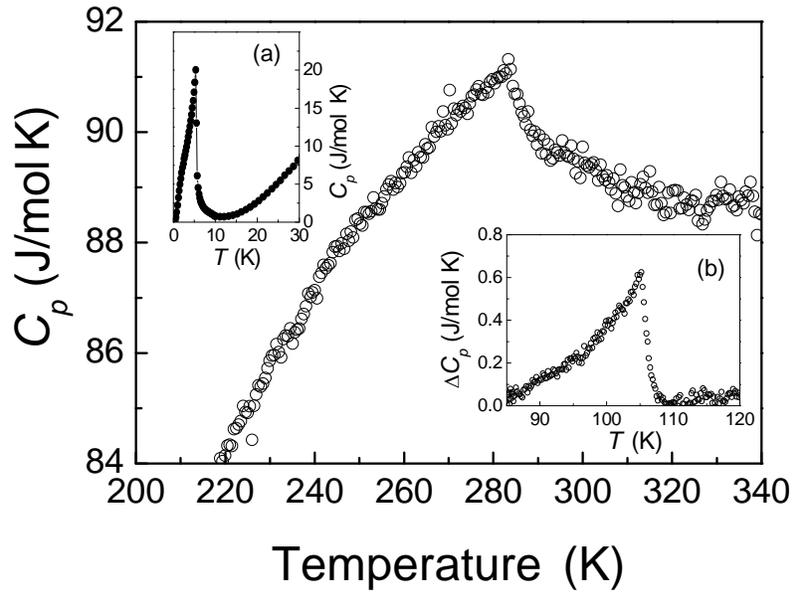

**Figure 3** Specific heat of ETO as a function of temperature in the temperature range around the phase transition. The insets show a) the low temperature region around $T_N$ with the lambda type anomaly in the specific heat; b) the specific heat anomaly $\Delta C_p$ of STO around the 105K transition. A background similar to the procedure suggested by Salje et al. [44] has been subtracted.

In summary, a phase transition has been predicted to exist in ETO, analogous to that in STO, wherein the oxygen octahedra tilt at the theoretical transition temperature $T_A \approx 298K$. The prediction has been confirmed experimentally by specific heat measurements which in comparison to STO clearly demonstrate its existence, however at $T_A=282(1)K$. In addition, also the antiferromagnetic transition is well detected as a λ-type anomaly at T=5.5K. While theory and experiment both reveal close analogies between STO and ETO, a distinctive difference in the dynamics exists, namely that ETO is more on the order / disorder side, whereas STO is in the displacive limit.

**Acknowledgement** One of us (ABH) wishes to acknowledge many useful discussions with S. Kamba.